\documentclass{JHEP3}
\usepackage{latexsym,amssymb,amsfonts,amsmath}
\usepackage{epsfig}
\newcommand{\la}{\langle}
\newcommand{\ra}{\rangle}
\newcommand{\nnn}{{\cal N}}

\newcommand{\eq}{\begin{equation}} 
\newcommand{\eqx}{\end{equation}}
\newcommand{\en}{\begin{enumerate}} 
\newcommand{\enx}{\end{enumerate}}
\newcommand{\ba}{\begin{eqnarray}} 
\newcommand{\ea}{\end{eqnarray}}
\newcommand{\bi}{\begin{itemize}} 
\newcommand{\ei}{\end{itemize}} 
\newcommand{\ii}{\item}
\newcommand{\s}{\varsigma}

\newcommand{\f}[2]{\frac{#1}{#2}}
\newcommand{\lra}{\longrightarrow}
\newcommand{\cor}[1]{\left\langle{#1}\right\rangle}

\newcommand{\ka}{\kappa}
\newcommand{\lam}{\lambda}
\newcommand{\xpr}{\vec b}
\newcommand{\q}{\vec q}

\newcommand{\zmx}[1]{z_{#1 \,{\rm max}}}
\newcommand{\alp}{\alpha'}
\newcommand{\al}{\alpha}
\newcommand{\bt}{\beta}
\renewcommand{\th}{\theta}

\newcommand{\gtt}{h_{\tau\tau}}
\newcommand{\gss}{h_{\sg\sg}}
\newcommand{\dl}{\delta}

\newcommand{\g}{\gamma}
\newcommand{\sg}{\sigma}

\newcommand{\lab}{\label}

\renewcommand{\Re}{{\rm Re}}
\renewcommand{\Im}{{\rm Im}}

\newcommand{\W}{{\cal W}}

\title{High Energy Bounds on Soft  $\nnn\!=\!4$ SYM Amplitudes from  AdS/CFT}

\author{M.~Giordano\\ Dipartimento di Fisica, Universit\`a di Pisa,\\
  Largo Pontecorvo 3, I-56127, Pisa, Italy, and \\ 
 Institut de Physique Th\'eorique  CEA-Saclay \\ F-91191
Gif-sur-Yvette Cedex, France \\ E-mail: \email{matteo.giordano@df.unipi.it,
matteo.giordano@cea.fr}}
\author{R.~Peschanski \\ Institut de Physique Th\'eorique  CEA-Saclay
  \\ F-91191 Gif-sur-Yvette Cedex, France \\ E-mail:
  \email{robi.peschanski@cea.fr}} 

\abstract{
Using  the AdS/CFT correspondence, we study the high-energy behavior
of colorless dipole elastic scattering amplitudes in {$\nnn\!=\!4$}
SYM gauge theory through the Wilson loop correlator formalism and
Euclidean to Minkowskian analytic continuation. The purely elastic 
behavior obtained at large impact-parameter $L$, through
duality from disconnected $AdS_5$ minimal surfaces beyond the
Gross-Ooguri transition point, is  combined  with  unitarity  and
analyticity constraints in the central region. In this way we obtain
an absolute 
bound on  the high-energy behavior of the forward scattering amplitude
 due to the graviton interaction between minimal
surfaces  in the bulk. The dominant ``Pomeron'' intercept is bounded
by $\al\le{11}/ 7$ using the AdS/CFT constraint of a weak
gravitational field in the bulk. Assuming the elastic eikonal
approximation in a larger impact-parameter range gives  $
4/3\le\al\le{11}/ 7.$ The actual intercept becomes $4/3$ if one assumes
the elastic eikonal approximation within its maximally allowed
range $L \gtrsim \exp{Y/3},$ where $Y$ is the total
rapidity. Subleading AdS/CFT 
contributions at large impact-parameter due to the other $d=10$
supergravity fields are obtained.  A divergence in the real part of
the tachyonic KK scalar is cured by analyticity but signals the need
for a theoretical completion of the AdS/CFT scheme.  
}

\preprint{
  Revised version\\
  April 2010} 

\begin{document}


\section{Introduction}
\label{intro}
From the point of view of the microscopic theory, the problem of high-energy
{\it soft}
($i.e.$, at small transverse momentum transfer) 
hadronic elastic amplitudes  has yet remained essentially
unsolved. Indeed, it involves the non perturbative regime of the
underlying Quantum Chromodynamics field theory (QCD) at strong
coupling, which, apart from lattice calculations, is beyond reach at the moment. However,
some fundamental properties have been known since a long time,
coming from the $S$-matrix formalism, and they are expected
to hold for a consistent quantum field theory. They  are {\it
  Unitarity,}  coming from the conservation of probabilities, and {\it
  Analyticity}. These properties, combined with  the existence of
a ``mass gap'' in the asymptotic particle spectrum,
lead to the celebrated Froissart bound for the total cross section \cite{Froissart,Froissart2,Froissart3},
which corresponds (up to logarithms) to an intercept not greater than $1$ 
for the leading Regge singularity, usually called the ``Pomeron''. 
To be more precise, in terms of the impact-parameter ($\vec b$) and rapidity ($Y$)
dependence of the partial elastic amplitude $a(Y,\vec b)$,  unitarity gives a
bound on $\Im \ a \le 2$ (in standard units), while confinement provides
a  bound  on the impact-parameter radius  $ L \propto Y.$  Both ingredients enter 
the derivation of the Froissart bound. 

Recently, a new tool for dealing with soft amplitudes has appeared, namely
 the Ga\-uge/Gravity duality, whose precise realization
has been first found~\cite{adscft,adscft2,adscft3,review} within the formalism of the
AdS/CFT correspondence. Generally speaking,  Gauge/Gravity duality  is
expected  to relate a strongly coupled  gauge field theory with a
``weakly coupled'' 5-dimensional supergravity limit of a
10-dimensional string theory. This raises the hope to find a solution
by mapping high-energy amplitudes into  supergravity by duality.  In
the realization of the AdS/CFT correspondence, the gauge theory is the
$\nnn=4$ supersymmetric Yang-Mills (SYM) gauge theory, which is a
conformal field theory and thus non-confining: the result should then
differ from what is found in a confining theory like QCD. However,
attacking the problem of soft amplitudes in this context may be a useful laboratory
for further developments in QCD. Indeed, the study of  soft high-energy
scattering amplitudes in  $\nnn=4$ SYM using the AdS/CFT
correspondence, and more generally Gauge/Gravity duality, has
attracted much attention in the 
literature~\cite{Jani,Jani1,Jani2,Brow0,Brow1,Brow2,Brow3,Brow4,Corn1,Corn2,Corn3,Corn4,Corn5,Tali1,Tali2,Muel,LP,Khar}. 

In the conformal case there is no mass gap, and thus the Froissart bound is
not expected to be valid for $\nnn =4$ SYM theory. However, unitarity and
analyticity are still expected to 
hold, and so it is interesting to examine the question of high-energy bounds
in this context. Indeed, in the perspective of applying the same
tools to gauge theories more similar to QCD, it is worthwhile to learn
more from the precise ``laboratory'' furnished by the AdS/CFT
correspondence. For this sake we shall use together unitarity,
analyticity and the AdS/CFT correspondence to give a precise account
of soft high-energy elastic amplitudes in the $\nnn=4$ supersymmetric
gauge theory. 

The difficult problem one is faced with when using the AdS/CFT correspondence
is that it applies  to the planar,  large-$N_c$ limit of
the gauge theory. This planar approximation corresponds
to purely elastic 
contributions since there is no particle production, and so it  does not take into
account  the  contribution of inelastic multiparticle channels, which are an
essential feature for determining  soft high-energy
amplitudes. However, the analytic continuation from Euclidean to
Minkowskian space may generate inelasticity for the scattering
amplitude (see Ref.~\cite{Jani1,Jani2}).

Our guideline is to circumvent this difficulty by combining the
knowledge one can obtain from AdS/CFT in the region of applicability
of the supergravity approximation, $i.e.$, the large impact-parameter
region where the amplitude is essentially elastic, with the
constraints coming from analyticity and  unitarity  which are expected
to hold for gauge field theories.
 
To be specific we shall consider the following ingredients:
\bi
\ii 
The r{\^o}le of massive quarks and antiquarks ($Q,\bar Q$)
in the AdS/CFT correspondence will be played, as in \cite{Wilson,Wilson2}, by
the massive $W$ bosons arising from breaking $U(N+1)\rightarrow
U(N)\times U(1),$ where one brane is considered away from the $N\to
\infty$ others. Hence, the role of  hadrons will be devoted to
``onia'' defined as linear combinations of $Q \bar Q$ colorless
``dipole'' states \cite{Muel1,Muel2,Muel3,Nave} of average transverse
size $\la{|\vec{R}|}\ra$, which sets the scale for the onium mass.  

\ii The dipole amplitudes will be defined through  correlators of
Wilson loops  in Euclidean space, in order to avoid the
complications 
of the Lorentzian AdS/CFT correspondence (see $e.g.$~\cite{lorads}). Hence, we will start from
the Euclidean formulation of the problem, and then  perform an
analytic continuation \cite{Megg,Megg2,Megg3,Megg4,Megg5,crossing,crossing2} to obtain the physical quantity in
Minkowski space~\cite{Nacht,Nachtr,DFK,LLCM1,BN}.  

\ii The AdS/CFT correspondence will allow to relate the calculation of
the Euclidean Wilson loop correlator to a {\it minimal surface} problem in
the Anti de Sitter bulk, which has been solved \cite{Jani} at large
impact-parameter distance $L$ using the knowledge of
(quasi-)disconnected minimal surfaces, connected by supergravity
fields propagating in the bulk. This is the solution of the minimal
surface problem  beyond the Gross-Ooguri transition point~\cite{Gross,Druk}.  

\ii After analytic continuation, this result will lead to an estimate
of  cross sections and elastic amplitudes at large impact
parameter. Combined with unitarity and analyticity to fix a bound in
the lower impact-parameter domain, it will lead to a determination of
new energy bounds on the forward elastic amplitudes, or, equivalently, 
on the total cross sections in $\nnn=4$ SYM . 
\ei

Note that we base our analysis on the use of minimal surface solutions
in AdS space with {\it Euclidean signature}~\cite{Jani,Jani1,Jani2}. 
More recently, the use of minimal surfaces in AdS space with {\it Minkowskian signature} 
for two- and many-body gluon
scattering has been developed (see~\cite{Alday,Alday2} and references
therein). In the present work we stick to the study of soft scattering
amplitudes for colorless states. 

The plan of the paper is as follows. In section 
\ref{Elas},
 we formulate the amplitudes in terms of Wilson loop correlators in
 Euclidean space, where the AdS/CFT correspondence is applied, and we
 define properly the analytic continuation to the physical Minkowski
 space. In section \ref{Mini}
we derive the formulation of  the AdS/CFT minimal surface solution
\cite{Jani} for the impact-parameter dependent amplitudes valid at
large impact-parameter (extended to unequal dipole sizes). In the following section \ref{strong0}, the
impact-parameter domain for the applicability of the AdS/CFT
correspondence is determined from the weak gravitational field constraint
in the AdS dual. Together with unitarity
constraints it allows us, in section \ref{forward}, to determine an absolute bound on the 
high-energy behavior of total cross sections and thus on the leading 
``Pomeron'' intercept of the forward elastic amplitude.  We briefly
discuss subleading contributions, among which the next-to-leading,
parity-odd one corresponds to the ``Odderon'' in $\nnn\!=\!4$ SYM
theory. In section \ref{concl}, we summarize our 
main results, compare them with existing studies and propose an
outlook on future related studies.

\section{Elastic amplitudes from Euclidean Wilson Loop correlators}
\label{Elas}
\FIGURE{
  \includegraphics[width=0.65\textwidth]{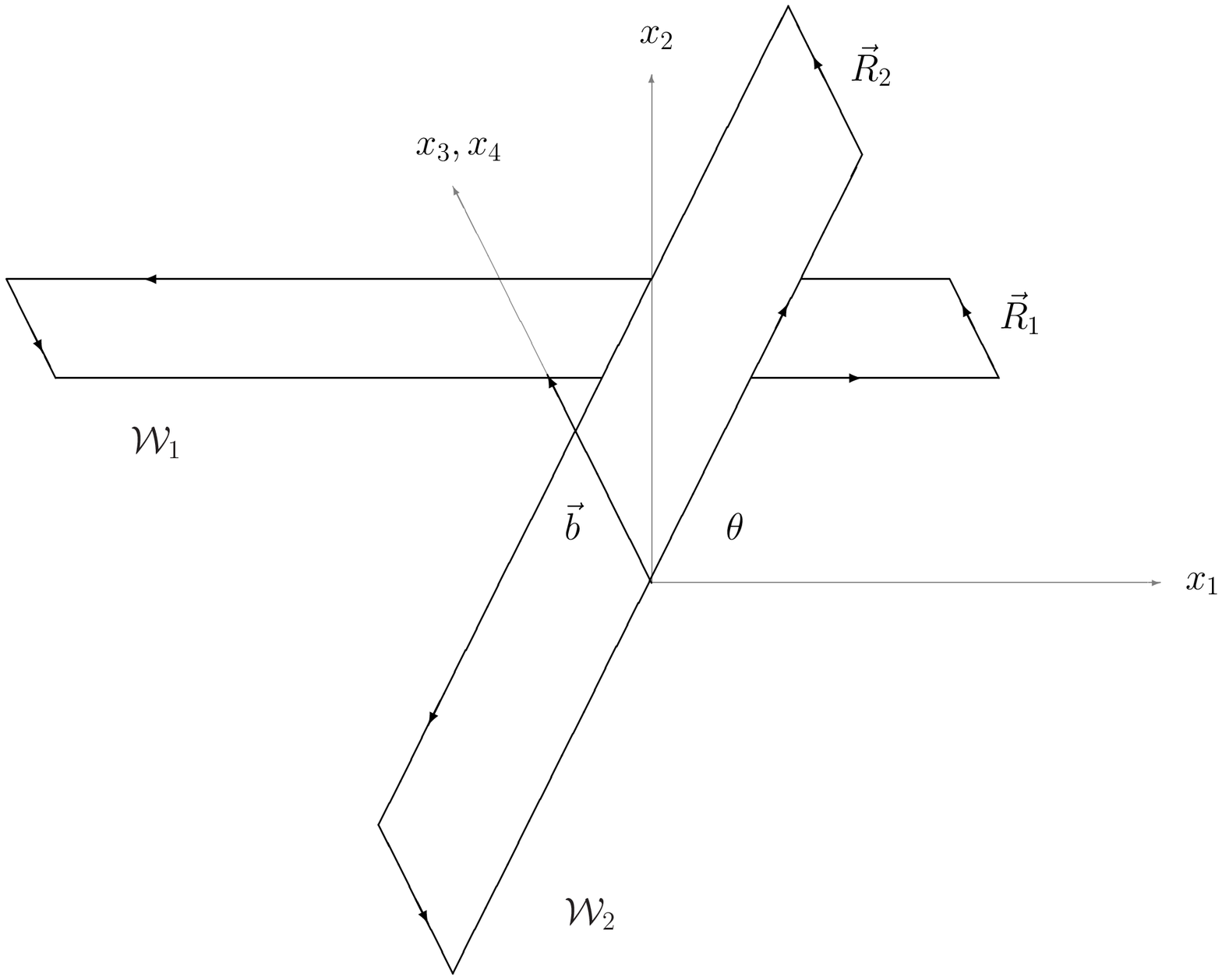}
  \caption{{\it Geometry of the Wilson loops in Euclidean space.} 
    The transverse kinematic variables ($\vec b,\vec R_{1,2})$ remain
    unchanged by the analytic continuation to Minkowski space, while $\th \to
    -i\chi$, see text.} 
\label{1}
}

In order to obtain information on high-energy elastic scattering
amplitudes at strong coupling, we will follow the approach
of~\cite{Jani} to evaluate them through the Gauge/Gravity
correspondence, making use of minimal surfaces in the gravity
bulk. The specific tool we will use is the AdS/CFT correspondence,
which allows to evaluate a certain Wilson-loop correlation function
in $\nnn =4$ SYM theory in Euclidean space through its gravitational dual, together
with analytic continuation into the 
physical Minkowski space, where this correlator corresponds to a
dipole-dipole elastic scattering amplitude. In this Section we
briefly describe the Wilson loop
formalism~\cite{Nacht,Nachtr,DFK,LLCM1,BN}, and the analytic continuation
required to relate Euclidean and Minkowskian quantities~\cite{Megg,Megg2,Megg3,Megg4,Megg5,crossing,crossing2}.

In the eikonal approximation, dipole-dipole elastic scattering
amplitudes in the high energy limit and at small momentum transfer
(the so-called {\it soft} high-energy regime) can be conveniently
expressed in terms of the normalized connected correlator ${\cal
  C}_M$ of Wilson {\em loops} in Minkowski space~\cite{Nachtr,DFK,LLCM1,BN}
\eq
\label{e.ampinit}
\begin{aligned}
{\cal A }(s,t;\vec{R}_1,\vec{R}_2)&= -2is \int
d^2\xpr\, e^{i\q\cdot\xpr}\, {\cal C}_M(\chi,\xpr;\vec{R}_1,\vec{R}_2)
\\ &\equiv -2is \int d^2\xpr\,  e^{i\q\cdot\xpr}
\cor{\f{\W_1\W_2}{\cor{\W_1}\cor{\W_2}}-1}, 
\end{aligned}
\eqx
where $t=-\vec{q}^{\,2}$, $\vec{q}$ being the transverse transferred
momentum (here and in the following we denote with
  $\vec{v}$ a two-dimensional vector), and the Wilson loops follow the classical  
straight-line trajectories for quarks (antiquarks, in parenthesis) \cite{Verl}:
\eq
\W_1\lra X^\mu=b^\mu+ u_1^\mu\tau\,(+r_1^\mu)\ ; \quad \quad \W_2\lra
X'^\mu=u_2^\mu\tau\, (+r_2^\mu)\ .
\label{traj}
\eqx
 Here $u^\mu_{1,2}$ are unit
time-like vectors along the directions of the momenta defining the dipole classical 
trajectories, and moreover $b^\mu=(0,0,\vec{b})$ and $r^\mu_{1,2} =
(0,0,\vec{R}_{1,2}).$ The loop contours are then closed at positive and negative infinite
proper-time $\tau$ in order to ensure gauge-invariance. 

The amplitude (\ref{e.ampinit}) corresponds to the
scattering of colorless quark-anti\-quark pairs with transverse
separation $R_i=|\vec{R}_i|.$  The scattering amplitude for two onium
states can then be
reconstructed from the dipole-dipole amplitude after folding with the
appropriate wave-functions for the onia,
\eq
{\cal A }^{(O)}(s,t)=
\int d^2\vec{R}_{1}\, d^2 \vec{R}_{2}\, |\psi_1(\vec{R}_{1})|^2
  |\psi_2(\vec{R}_{2})|^2\, {\cal A }(s,t;\vec{R}_1,\vec{R}_2)\ .
\eqx The mass of an onium state with wave
function $\psi$ is expected to be of the order of the inverse of its
average radius, $m\sim [\int d^2\vec{R}\, |\psi(\vec{R})|^2\,
  R]^{-1}\equiv\la R \ra^{-1}$.

The geometrical parameters of the configuration can be related to the
energy scales by the relation
\eq
\cosh \chi\equiv \f{1}{\sqrt{1-v^2}}=\f{s}{2m_1m_2}-\f {m_1^2+m_2^2}{2m_1m_2}\equiv \s -\f {m_1^2+m_2^2}{2m_1m_2},
\lab{chi}
\eqx
where $\chi=\f{1}{2}\log \f{1+v}{1-v}$
is the hyperbolic angle 
(rapidity) between the trajectories of the dipoles,
$v$ is their relative velocity, $Y=\log\s$ the total rapidity and $m_{1,2}$ the masses of the  onia.

The formulation of the Gauge/Gravity correspondence that we want to use relates the Wilson loop correlator in formula 
\eqref{e.ampinit} to the solution of a minimal surface problem in the bulk of
the dual 5-dimensional space. This rule has been applied for scattering amplitudes to various
dual geometries~\cite{Jani,Jani1,Jani2}. The determination of the minimal surface in
a given gravity background is in general a difficult mathematical
problem. 
However, there are known situations where
the minimal surfaces can be obtained analytically, in particular for the 
case of the AdS/CFT correspondence between the $\nnn=4$ SYM gauge theory and 
the $AdS_5$ geometry. In order to avoid the complications related to
the Minkowskian signature~\cite{lorads}, it is convenient to
exploit the Euclidean version of the correspondence, and then to 
reconstruct the relevant correlation function ${\cal C}_M$ from its
Euclidean counterpart ${\cal C}_E$ by means of analytic
continuation~\cite{Megg,Megg2,Megg3,Megg4,Megg5}. The Euclidean approach has already been
employed in the study of high-energy soft scattering amplitudes by
means of non perturbative
techniques~\cite{Jani,Jani1,Jani2,ILM,LLCM2}, including numerical 
lattice calculations~\cite{Latt,Latt2}.

The Euclidean normalized connected correlation
function is defined as
\ba
  \label{eq:sugraexch0}
  {\cal C}_E(\theta,\xpr;\vec{R}_1,\vec{R}_2) &\equiv & \frac{\la \W_1
    \W_2 \ra}{\la \W_1 \ra \la \W_2 \ra} - 1\ , 
\ea
where $\W_i$ are now Euclidean Wilson loops evaluated along the
straight-line paths $\W_1\lra X^\mu=b^\mu+u_1^\mu\tau\,(+r_1^\mu)$ and $\W_2\lra
X'^\mu=u_2^\mu\tau\, (+r_2^\mu)$, closed at infinite proper
time, see Fig.~\ref{1}. The variables $b$ and $r_i$ are the same defined above in the Minkowskian
case (we take Euclidean time to be the first coordinate to
  keep the notation close to the Minkowskian case, see \eqref {traj}).
Here $u_1$ and $u_2$ are unit vectors forming an angle $\theta$
in Euclidean space.

The physical correlation function ${\cal C}_M$ in Minkowski space is
obtained by means of the analytic continuation~\cite{Megg,Megg2,Megg3,Megg4,Megg5}
\eq
\label{e.analcont}
\th \lra -i\chi\mathop\sim_{s\to\infty} -i \log\s\ 
\eqx
from the Euclidean correlator ${\cal C}_E$. To be more precise, the
two quantities are related through the analytic continuation relation
\eq
\label{e.ancont}
{\cal C}_M(\chi,\vec{b};\vec{R}_1,\vec{R}_2) =
{\cal C}_E(-i\chi,\vec{b};\vec{R}_1,\vec{R}_2),\quad \chi\in\mathbb{R}^+ \ ,
\eqx
where the analytic continuation of ${\cal C}_E$ is performed starting from 
the interval $\theta\in (0,\pi)$ for the Euclidean angle (the
restriction to positive values of $\chi$ and to $\theta\in (0,\pi)$
does not imply any loss of information, due to the symmetries of the
two theories).

It is worth mentioning that combining the analytic-continuation
relation \eqref{e.ancont} with the symmetries of the Euclidean theory
one can derive non-trivial crossing-symmetry relations for the
Minkowskian loop-loop correlator~\cite{crossing,crossing2},
\eq
{\mathcal{C}}_M(i\pi-\chi,\vec{b};\vec{R}_1,\vec{R}_2)
=\mathcal{C}_M(\chi,\vec{b};\vec{R}_1,-\vec{R}_2)
=\mathcal{C}_M(\chi,\vec{b};-\vec{R}_1,\vec{R}_2), \, \chi\in\mathbb{R}^+\ .
\eqx
These relations allow to decompose the amplitudes in
crossing-symmetric and crossing-antisymmetric components as follows,
\eq
\begin{aligned}
{\cal A}^{(\pm)}(s,t;\vec{R}_1,\vec{R}_2) &= -i2s \int
d^2\xpr\ e^{i\q\cdot\xpr}\,{\cal C}_M^{(\pm)}(\chi,\xpr;\vec{R}_1,\vec{R}_2)\ ,\\
{\cal C}_M^{(\pm)}(\chi,\xpr;\vec{R}_1,\vec{R}_2) &\equiv \f{1}{2}\left({\cal
  C}_M(\chi,\xpr;\vec{R}_1,\vec{R}_2) \pm {\cal
  C}_M(i\pi-\chi,\xpr;\vec{R}_1,\vec{R}_2)\right)\ . 
\label{Sign}\end{aligned}
\eqx

In the next Section we will describe how the Wilson-loop correlator
is related to minimal surface solutions via the AdS/CFT
correspondence. The result will be used to estimate the high-energy
elastic dipole-dipole scattering amplitude in $\nnn=4$ SYM gauge
theory, in the appropriate kinematic range where the planar
(large-$N_c$) approximation is expected to be valid.

\section{Wilson loop correlators from AdS/CFT}
\label{Mini}
\FIGURE[t]{
\centering
\includegraphics[width=0.8\textwidth]{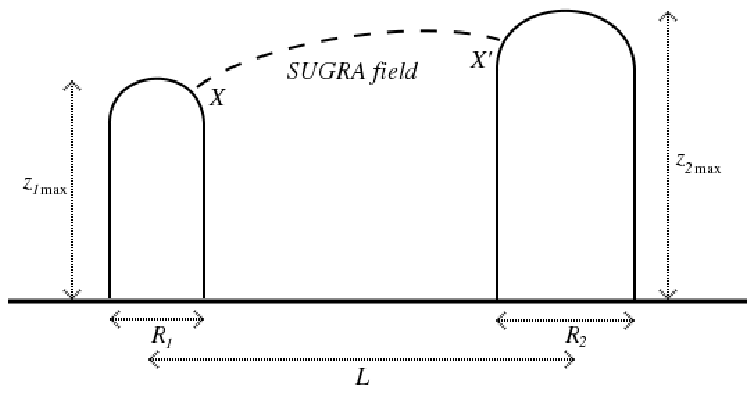}
\caption{{\it Correlation function of Wilson loops from AdS/CFT.} The correlation function
is calculated through
the exchange of bulk supergravity fields between minimal surfaces attached to each individual
Wilson loop. Here $\zmx{i}$ are the depths of the surfaces in the 5th
dimension of AdS. The other kinematic
notations are 
as in Fig.~\ref{1}.}
\label{2}
}

Within the AdS/CFT correspondence, the correlators of Wilson loops
in the gauge theory, such as those of 
Eq.~\eqref{e.ampinit}, are related to a minimal surface in the bulk of
$AdS_5$ having as boundaries the two Wilson loops, which corresponds to minimizing the Nambu-Goto action. The 
{\it analytic} solution of a minimal surface problem in Euclidean space (the so-called Plateau 
problem) is in general a highly nontrivial mathematical issue, and it
is even more so in a non flat metric such as $AdS_5$. For our purpose,
an analytic solution is required in order to adequately perform the
Euclidean-to-Minkowskian analytic continuation.  

Our guiding line, following Ref.~\cite{Jani}, is that the solution simplifies
provided the impact parameter distance $L\equiv \vert \vec b \vert \gg
R_1,R_2$. 
Indeed, when $L \lesssim R_1,R_2 $  there exists a connected minimal
surface with the sum of the two loops as its disjoint boundary (see
$e.g.$~\cite{Zarembo}), although its explicit expression is 
difficult to obtain. However, when $L\gg R_1,R_2$ the minimal surface
has two independent, (quasi-)disconnected components: in order to calculate
the correlator one then exploits, as in Ref.~\cite{MaldCor}, the
explicit solutions corresponding to the two loops connected by the
classical supergravity interaction, $i.e.$, by the exchange  (see
Fig.~\ref{2}) between them of the lightest fields of the $AdS_5$
supergravity, namely the graviton, the anti-symmetric tensor and the
dilaton, which are massless, and the tachyonic Kaluza-Klein (KK)
scalar mode. This is the case we consider here, using the large-$L$ 
behavior of the dipole-dipole impact-parameter amplitude evaluated
in~\cite{Jani}, with a generalization to unequal dipole sizes.
As we have already pointed out, we start from the Euclidean
formulation of the problem, and so we consider minimal surfaces in
$AdS_5$ with Euclidean metric in order to obtain the Wilson-loop
correlation function in $\nnn\!=\!4$ SYM theory in Euclidean space.

Let us recall the main results of~\cite{Jani} and describe briefly 
how one determines 
the leading dependence on the impact parameter $L$ and on the rapidity
$\chi$ of the various phase shifts (corresponding to the various
exchanges of supergravity fields), together with the dependence on the
size of the dipoles. 
For $L\gg R_1,R_2$, and in the weak gravitational field domain, the
Euclidean normalized connected correlation function has the form  
\eq
\begin{aligned}
  \label{eq:sugraexch}
  {\cal C}_E  & 
  =  \exp{\left(\sum_\psi \tilde{\dl}_\psi\right)} -1\ , \\
  \tilde{\dl}_\psi &\equiv  \frac{1}{4\pi^2 \alpha'{}^2} 4 \int d\tau_1 d\tau_2 \,
  \frac{dz_1}{z_{1x}}\frac{dz_2}{z_{2x}} \frac{\delta S_{NG}}{\delta \psi}(\tau_1,z_1)
\  G_\psi(X,X') \frac{\delta S_{NG}}{\delta \psi}(\tau_2,z_2)\ ,
\end{aligned}
\eqx
where $S_{NG}$ is the Nambu-Goto action, $\alpha'=1/\sqrt{4\pi g_sN_c}$ and  $g_s$ is the string coupling.
Here $\frac{\delta S_{NG}}{\delta \psi}$ is the coupling of the
world-sheet minimal surfaces attached to the two Wilson loops to the supergravity field
$\psi$. Moreover, $\tau_i$ is the proper time on 
world-sheet $i=(1,2)$, and $z_{1,2}$ are the fifth coordinates of points $X$, $X'$
in $AdS_5$, namely
\eq
  \label{eq:coord}
  X=\left(u_1^\mu\tau_1 + x_1^\mu + b^\mu,z_1\right), \quad
  X'=\left(u_2^\mu\tau_2 + x_2^\mu,z_2\right)\ .
\eqx
For the relevant four-dimensional vectors we use the notation
\eq
\begin{aligned}
 u_1^\mu &= (1,0,\vec{0}), & u_2^\mu =
  (\cos\theta,-\sin\theta,\vec{0})\ ,\\
x_{i}^\mu &= \sigma_{i}(z_i) \f{r_i^\mu}{R_i}, & \sigma_i\in[0,R_i],\quad i=1,2\ ,
\end{aligned}
\eqx
where $\sigma_{i}(z_i)$ is determined by inverting the solution of the
minimal surface equation $z_i=z_i(\sigma_i)$. The derivatives $z_{ix}\equiv
\f{\partial z_{i}}{\partial \sigma_{i}}$ are given by~\cite{Wilson,Wilson2} 
\ba
z_{ix}=\left(\frac{z_{i{\,\rm
      max}}}{z_i}\right)^2\sqrt{1-\left(\frac{z_i}{z_{i {\,\rm max}}}\right)^4}, 
\quad z_{i\,\rm{max}}={R}_{i}\,\frac{[\Gamma(1/4)]^2}{(2\pi)^{3/2}}\ .
\label{minimal}
\ea 
In Eq.~\eqref{eq:sugraexch}, $G_\psi(X,X')$ is the Green function relevant to the exchange of field
$\psi$, which depends only on invariant bitensors and scalar
functions~\cite{bitensor} of the AdS invariant
\begin{equation}
  \label{eq:AdSinv}
  u = \frac{(z_1-z_2)^2 + \sum_{j=1}^4 (X_j-X'_j)^2}{2z_1z_2}\ .
\end{equation}
By the change of coordinates $v_+=\tau_2\sin\theta$,
$v_-=\tau_1-\tau_2\cos\theta$, one allows~\cite{Jani}
the dependence on $\theta$ to drop from $G_\psi(X,X')$, so that it can
be read off directly from the couplings. One is then able to
isolate the leading dependence on $L$ and $R_{i}$ by performing the
rescaling $\zeta_i=z_i/z_{i\,{\rm max}}$, with $\zeta_i\in[0,1]$, and
$\rho_\pm=v_\pm/L$: we obtain a factor $L^2/z_{1\,{\rm max}}z_{2\,{\rm
    max}}$ in front of the integrals, and we find the leading term in $u$ to
be ${L^2}/z_{1\,{\rm max}} z_{2\,{\rm max}}$ times a function of
the new integration variables. 
Working out the Green functions and the couplings 
corresponding to the exchange of the various supergravity fields,
and performing the remaining integrals, it is found that the leading
dependence (the ``leading'' term in $\theta$ is understood as
  the leading term in $\chi$ after analytic continuation, see below)  on
$\theta$, $L$ and  ${R}_{i}$ for the various 
terms of \eqref{eq:sugraexch} is the following\footnote{The
  $L$-dependence of the dilaton contribution has been corrected with
  respect to the misprinted one reported in~\cite{Jani}.}: 
\eq
\begin{alignedat}{3}
\label{eq:phaseshifts}
 \tilde{\delta}_S \ \ &= \ \   \kappa_S\ \frac{1}{\sin\theta}
  \left(\frac{{R}_{1}{R}_{2}}{L^2}\right)&& \ \ \equiv \ \   a_S
  \ \frac{1}{\sin\theta} && \ \ \quad  \text{ (KK scalar)}\ ;\\
 \tilde{\delta}_D \ \ &= \ \   \kappa_D\  \frac{1}{\sin\theta}
  \left(\frac{{R}_{1}{R}_{2}}{L^2}\right)^3&& \ \ \equiv \ \   a_D
  \ \frac{1}{\sin\theta} && \ \  \quad\text{ (dilaton)}\ ;\\
 \tilde{\delta}_B \ \ &= \ \   \kappa_B\  
\frac{\cos\theta}{\sin\theta}\left(\frac{{R}_{1}{R}_{2}}
{L^2}\right)^2&& \ \ \equiv \ \   a_B
  \ \frac{\cos\theta}{\sin\theta} && \ \ \quad \text{ (antisymmetric tensor)}\ ;\\
 \tilde{\delta}_G \ \ &= \ \   \kappa_G\ 
\frac{(\cos\theta)^2}{\sin\theta}\left(\frac{{R}_{1}{R}_{2}}
  {L^2}\right)^3&& \ \ \equiv \ \   a_G
  \ \frac{(\cos\theta)^2}{\sin\theta} && \ \ \quad \text{ (graviton)}\ ,
\end{alignedat}
\eqx
factorizing
explicitly the angular dependence from the rest. In \eqref{eq:phaseshifts} 
we keep track of the different dipole sizes.
To be complete, the factors $\kappa_{\psi}$ for each
supergravity field are given by (see formulas (34,38,49,58)
of~\cite{Jani})
\ba
\ka_S=\f{g_s}{2N_c} \f{10}{\pi^2}\ ,\quad \ka_D=\f{g_s}{2N_c}\ \f 3{16}
\left[\f{\Gamma(1/4)}{2\pi}\right]^8\ ,\quad \ka_B = 
 {\rm const}\cdot\f{g_s}{2N_c}\ ,
\label{kappas}
\ea
where the last constant is yet to be determined, and for the graviton field
\ba
  \kappa_G =
\frac{g_s}{2N_c}\frac{15}{2}\left[\frac{\,[\Gamma(1/4)]^2}{4\pi}\right]^4\ .
\label{graviton}
\ea 
The string coupling $g_s$ is  related to the gauge
theory coupling by $g_{\rm YM}^2 =2\pi g_s$.
Performing now the analytic continuation $\theta\to -i\chi$, leading to the 
phase shifts $i\delta_\psi\equiv{\tilde\delta_\psi}(\theta\to-i\chi)$, one
finally obtains for the Minkowskian correlation function
\eq
  {\cal C}_M = \exp\left(i\sum_\psi \delta_\psi\right)-1\ ,
\lab{CM}
\eqx
where $\delta_\psi$ are given by 
\begin{equation}
\label{eq:phaseshiftsM}
  \delta_S = a_S\,\f{1}{\sinh\chi},\quad   \delta_D = a_D\,\f{1}{\sinh\chi},\quad
  \delta_B = a_B\,\f{\cosh\chi}{\sinh\chi},\quad   \delta_G = a_G\,\f{(\cosh\chi)^2}{\sinh\chi}\ .
\end{equation}
Since the functions $a_\psi$ depend only on the combination
$\frac{L^2}{R_1 R_2}$ of the moduli of the impact parameter and of
the dipole sizes $R_{i}$, we will sometimes write  ${\cal C}_M\equiv {\cal C}_M(\chi,L;R_1,R_2)$.
We notice that under crossing, $i.e.$, under $\chi\to i\pi-\chi$, the
phases $\delta_S$, $\delta_D$ and $\delta_G$ are symmetric, while
$\delta_B$ is antisymmetric. One has then for the definite-signature
quantities in \eqref{Sign} the expressions
\begin{equation}
\begin{aligned}
  {\cal C}_M^{(+)}&= \cos(\delta_B)\exp\left[i\left(\dl_S + \dl_D + \dl_G\right)\right]-1\ ,\\
  {\cal C}_M^{(-)}&= i\sin(\delta_B)\exp\left[i\left(\dl_S + \dl_D + \dl_G\right)\right]\ .
\end{aligned}
\end{equation}


\section{AdS/CFT domain of validity}
\label{strong0}
\subsection{The weak field constraint}
The range of validity of the calculations above is determined by
requiring~\cite{Jani} that the effect of the gravitational perturbation $\dl
G_{tt}$ generated by each of the string world-sheets on the other one
is smaller than the background metric $G_{tt}$, therefore ensuring that
one is actually 
working in the weak-field limit. 
Considering the effect of world-sheet 2 on world-sheet 1 (see Fig.~\ref{2})
the strongest constraint is obtained from the evaluation of the maximal gravitational
field produced at the point $\tau_1=0,$ where the distance between the loops is minimal. The weak gravitational field
requirement reads
\eq
\label{e.gttc}
 \f{\dl G_{tt}}{G_{tt}}\ll 1, \quad G_{tt}\equiv\f{1}{z_1^2}\ ,
\eqx
where $G_{tt}$ is the background metric term coming from the $AdS_5$
Fefferman-Graham parameterization. In order to find the explicit expression 
of condition \eqref{e.gttc}, we note that the phase shift in
\eqref{eq:sugraexch} can be also interpreted as 
an integral over the string world-sheet 1 (at $\th=0$) of the corresponding supergravity field $\psi(X')$
produced by the other, tilted world-sheet 2 (at $\th$), namely 
\eq
\label{e.field}
\delta \psi (X(\tau_1,z_1))= \f{1}{2\pi \alp}  \int 2 d\tau_2 \f{dz_2}{z_{2_x}} 
\ G_\psi(X,X')\ \frac{\delta S_{NG}}{\delta \psi}(\tau_2,z_2)
\ . 
\eqx
Applying this generic equation to the specific dominant graviton contribution,
one writes the graviton coupling by expanding the Nambu-Goto
action, namely
\eq
\label{e.gravcp}
\sqrt{h}\left[ \f{\dl\gtt}{2\gtt} +\f{\dl\gss}{2\gss} \right] \sim
\f{\zmx{2}^2}{2{z_2}^2}\,\dl\gtt \sim (\cos\th)^2 \, \dl G_{tt}\ ,
\eqx
where $h_{ab}$ is the induced metric on world-sheet 1, and we have retained only the dominant part 
of the field after analytic continuation $\th \to -i\chi$. The 
field produced at the point $\tau_1=0$ by the second (tilted)
world-sheet is then given by 
\eq
\f{1}{2\pi\al'}\int 2 d\tau_2 \f{dz_2}{z_{2x}} \f{\zmx{2}^2}{2z_2^2} \cor{\dl\gtt^\th
\dl\gtt^{\th=0}}\ ,
\eqx
where the correlation function
\eq
\cor{\dl\gtt^\th \dl\gtt^{\th=0}} =\f{1}{z_1^2z_2^2
}\left[2G(u)(\cos\th)^2
+H(u) \right]\ 
\eqx
is given \cite{Dokk} in terms of the functions
$G(u)\sim (3/32\pi^2)\cdot 1/u^4$ and $H(u)\sim (-1/48\pi^2) \cdot
1/u^2$ for large $u,$ where $u$ is the AdS invariant \eqref{eq:AdSinv}.
Using $u(\tau_1=0)\sim(L^2+\tau_2^2)/(2z_1 z_2)$ one gets
\eq
\dl G_{tt} \propto (\cos\th)^2\,\int\f{dz_2}{z_{2x}}
\f{z_{2\,{\rm max}}^2}{z_2^2}\f{1}{z_1^2 z_2^2} 
 \int_{-\infty}^\infty d\tau_2\,\f{z_1^4 z_2^4}{(L^2+\tau_2^2)^4} \ ;
\eqx
one then performs the rescaling $\tau_2\to \tau_2/L,$ $z_2\to z_2/z_{2\,{\rm max}},$
with $z_{2\,{\rm max}} \propto R_2$, and finds the constraint
\eq
\f{\dl G_{tt}}{G_{tt}} \propto \f{z_1^4 R_2^3}{L^7}(\cos\th)^2 \ll 1\ .
\eqx
This constraint is most restrictive when evaluated at 
$z_1= z_{1\,{\rm max}}\propto R_1$, which
is as far as the string world-sheet extends into the $5${th} dimension
of $AdS_5,$ see Fig.~\ref{2}.
Performing the analytic continuation $\theta\to
-i\chi$, and interchanging the r\^oles of the two
world-sheets by switching the subscripts 1 and 2 in the
results above in order to get the maximal constraint, one finally obtains  
\eq
\label{e.constraint}
  \left(\frac{L^2}{{R}_{1}{R}_{2}}\right)^{\f{7}{2}} {\rm
     min}\left(\sqrt{\frac{{R}_{1}}{{R}_{2}}},\sqrt{\frac{{R}_{2}}{{R}_{1}}}\,\right) 
     \gg \ \s^2 \quad\! \Rightarrow \quad\! {L^2} \gg L_{max}^2 \equiv \f {{{R}_{1}{R}_{2}}\s^{\f47}}{\left[{\rm
     min}\left(\sqrt{\frac{{R}_{1}}{{R}_{2}}},\sqrt{\frac{{R}_{2}}{{R}_{1}}}\,\right)\right]^{\f 27}}\ ,
\eqx
and in this region the phase $\dl_G$ is actually small, as it should
be. It is interesting to remark  
that the smaller the dipole size, the larger the impact-parameter
region where the weak field approximation is valid, as we may expect from physical
intuition. Moreover, considering the possibility of an energy-dependent dipole
size inside a target, $e.g.$ due to high density effects\footnote{Such an effect could mimic 
the high partonic density effects in QCD
deep-inelastic scattering for which the dipole size $R$ in a high density
target would be of order $R\sim \s^{-\f12 \lam}$ with $\lam \sim .3$, see $e.g.$~\cite{Mueller}.}, the size 
dependence may even strengthen the cross section bound (see next section \ref{strong}) through
a weaker energy dependence of the impact-parameter limit $L_{max}$.  
Let us for instance assume an energy 
dependence $R_1^2\sim R_2^2\s^{-\lam}$ of the smaller dipole size: in this case 
one would find a weaker constraint on $L$, 
\eq
\lab{Satur}
\left(\f {L}{R_2}\right)^2\gg\ \s^{\f47-\f{3}{7} \lam}\ .
\eqx

\subsection{The elastic eikonal hypothesis}
\label{strong}

From expression \eqref{e.ampinit} one can determine the
impact-parameter partial amplitude $a(\chi,\xpr)$ corresponding to the
dipole-dipole elastic  amplitude ${\cal A}$, $i.e.$ (suppressing the
sizes of the dipoles) $a(\chi,\xpr) = -i{\cal
  C}_M(\chi,\xpr)$. 
In the large-$L$ region, following section \ref{Mini}, the AdS/CFT contribution reads
\eq
\label{e.impact}
a_{tail}(\chi,\xpr) =  i\left[1-\exp\left(i\sum_\psi
\dl_\psi\right)\right]\ ,
\eqx
with the phase shifts specified by \eqref{eq:phaseshiftsM}.
This expression can be trusted as long as the solution for the
minimal surface problem is disconnected, and above all, as
remarked in the preceding subsection, as long as the weak gravitational field
constraint \eqref{e.constraint} is satisfied. Note that since expression
\eqref{e.impact} verifies
the relation  
\ba
\Im  \ a_{tail}(\chi,\xpr)= \f12 \left\vert a_{tail}(\chi,\xpr)\right\vert^2\ ,
\lab{Unel}
\ea 
it corresponds to a purely elastic amplitude, in agreement with the planar limit implied
by the AdS/CFT correspondence.

The result  \eqref{e.constraint} calls for an important comment: it expresses a stringent constraint on the
impact-parameter range due to the weak gravitational field condition required
in applying the AdS/CFT correspondence. Let us now add to the discussion  
a possible extension of the results, obtained adopting the $S$-matrix point-of-view, but not 
{\it a priori} borne out by the dual AdS/CFT picture.

From the $S$-matrix point-of-view, the exponential  form of \eqref{e.impact}
(see also \eqref{CM}) is typical of  a resummation of non-interacting
($i.e.$, independent) colorless exchanges (on the gauge theory side) which can be taken into 
account in order to possibly enlarge the domain of validity of \eqref{e.impact}. This amounts 
to assume the validity of the eikonal approximation for a purely elastic scattering amplitude 
(see \cite{Itzy,Itzy2,Itzy3} for the eikonal approximation in QED; for QCD see $e.g.$~\cite{Itzy4}). 
However, in the framework of the microscopic theory, $i.e.$, the 4-dimensional gauge theory at strong coupling,
there is no  rigorous theoretical derivation of the eikonal formula. 
In fact, as a result of our previous analysis, we do not expect this independent 
resummation to be valid from the AdS/CFT correspondence point-of-view, since
a strong gravitational field  in the bulk near the relevant minimal surfaces
is expected to be the seed of graviton self-interactions, which would  spoil the independent 
emission of the gravitational eikonal formalism.

Yet, for completion, let us suppose that the eikonal formalism
for the elastic amplitude may be extended in some larger phase-space region and thus examine, 
from the empirical $S$-matrix  point-of-view, 
whether and  down to which value of the impact-parameter separation the formula \eqref{e.impact}
could be used beyond the constraint  \eqref{e.constraint}.
As a first step beyond our AdS/CFT correspondence result, one could infer from an $S$-matrix 
model formulation that the amplitude 
\eqref{e.impact}
is reliable as long as
the dominant graviton-induced phase shift $\dl_G$ is small. Following formulas 
(\ref{eq:phaseshiftsM}), this means that
$\frac{L^2}{{R}_{1}{R}_{2}}
  \gg \left(\kappa_G\f{(\cosh\chi)^2}{ 
\sinh\chi}\right)^{\f13}\sim \left(\kappa_G\s\right)^{\f13}.$ In fact  the minimal
 impact-parameter value for the eikonal formula \eqref{e.impact} to be physically sensible
 from the 4-dimensional point-of-view is more precisely
 \eq
{L^2}> L^2_{min}\equiv {{R}_{1}{R}_{2}}
\left(\f {\kappa_G}\pi\ \s\right)^{\f 13}
\lab{limit}
\eqx
 requiring the phase shift
 $\dl_G\le \pi ,$ 
see Eq.~\eqref{eq:phaseshiftsM}. This extreme minimal bound ensures that
$\Im\,a_{tail}(\chi,\xpr)$ be not oscillating with $L$: since it is just proportional, 
$via$ the optical theorem, to the $L$-dependent partial cross section, a non-oscillating 
behavior is expected. Indeed, it is reasonable to expect that more
and more inelastic channels would open up when going from the peripheral
to the central impact-parameter domain.  

From the dual gravitational point-of-view, the problem seems  severe.
Studies of the gravitational eikonal approximation already exist in the 
literature~\cite{Brow1,Corn1,Corn2,Corn3,Corn4,Corn5,LP}; 
however, the precise question which is relevant in our case is beyond which value
in the large impact-parameter range the eikonal expression \eqref{e.impact}
for a {\it purely elastic} amplitude is expected to be valid.  
This is equivalent to ask up to what impact-parameter distance it is mainly the exchange of 
independent gravitons in the bulk which builds the whole amplitude.
Using our minimal surface approach, we see that the limitation to a weak 
field approximation for the gravitational field at the tip of the minimal
surfaces gives a stronger constraint than the one \eqref{limit} coming from
the $S$-matrix model point-of-view.

\subsection{Characteristic impact-parameter scales}

Let us consider  a range of validity
of  \eqref{e.impact} varying from its AdS/CFT value defined
by \eqref{e.constraint} to its  maximal $S$-matrix model extension 
 \eqref{limit}. We are lead to define  a characteristic distance
$L_{tail}$ such that for $L=|\xpr|>L_{tail}(s)$ the impact-parameter
scattering amplitude is given by Eq.~\eqref{e.impact}.
One can then divide the whole impact-parameter space into 
a $tail$ region ($L>L_{tail}$), and a $core$ region ($L<L_{tail}$)
where inelastic channels are supposed to open up.
More specifically, the following regions are identified. 
\begin{enumerate}
\item At  large distances $L > L_{max},$ whose  exact expression is given by
  \eqref{e.constraint}, the gravitational field in the bulk is weak enough,
  and the contribution of the disconnected minimal surface gives a rigorous
  holographic determination of the impact-parameter tail of the scattering amplitude.
\item At moderately large distances $L_{min}<L<L_{max}$, where $L_{min}$
has been defined in \eqref{limit}, the strong gravitational field is expected
to generate a non zero $\Im\,\dl_G$ leading to inelastic contributions on the
gauge theory side, and hence to  $\Im  \ a_{tail}(\chi,\xpr)> \f12 \vert
 a_{tail}(\chi,\xpr)\vert^2$ contrary to \eqref{Unel}. 
The minimal surface is still disconnected but the gravitational field
begins to become strong in some relevant region in the bulk.
Nevertheless, for the sake of completeness, we will investigate what happens
assuming the validity of the elastic eikonal expression up to $L_{tail}$, lying somewhere 
in the range $L_{min}\le L_{tail}\le L_{max}$.
\item For $L_{connect}<L<L_{min}$ the elastic eikonal expression
  \eqref{e.impact} is no more reliable, even from the $S$-matrix
  point-of-view. An eikonal formula may still be valid with an  imaginary
  contribution to the phase shifts but it cannot be obtained through  the weak
  gravity regime of the AdS/CFT correspondence, even if the minimal  surface
  is still made of  disconnected surfaces joined by interacting 
  fields. 
\item Finally, for even smaller distances $L\le L_{connect}$ the Gross-Ooguri
 transition takes place, and the minimal surface solution becomes connected.
  In this region, the AdS/CFT description goes beyond the interaction mediated by
  supergravity fields.
\end{enumerate}

Region 1 and possibly part of region 2 constitute the impact-parameter {\it tail} 
region, while the regions 3 and 4 constitute the central impact-parameter {\it core} 
region. To incorporate these regions in our analysis we need to use information
coming from a source other than the AdS/CFT correspondence: in practice, we
will use the unitarity constraint 
\eq
\Im  \ a(\chi,\xpr)\le 2\ .
\label{Unit}
\eqx
Since we know precisely $a(\chi,\xpr)$ only in the $tail$
region, we are able to determine only part of the full scattering 
amplitude, $i.e.$, the large impact-parameter contribution ${\cal A }_{tail}$,
\eq
{\cal A }\equiv {\cal A }_{core}+{\cal A }_{tail}\ ;\quad{\cal A }_{tail}(s,t;\vec{R}_1,\vec{R}_2) = 2is \int_{L\ge L_{tail}}
d^2\xpr\ e^{i\q\cdot\xpr}\,\left[1-e^{\left(i\sum_\psi \dl_\psi\right)}\right]\ ,
\label{Tail}
\eqx
where ${\cal A }_{core}$
 will be  constrained by the unitarity bound \eqref{Unit}.
Exploiting this expression and \eqref{Unit}, we will be able to set a lower and
an upper bound on the large-$s$ behavior of the full amplitude
depending on the $s$-dependence of $L_{tail}$.

An important addendum to this discussion is related to its modification due to
energy-dependent dipole sizes, as in \eqref{Satur}. Indeed, sticking to the 
rigorous result coming from the AdS/CFT correspondence in the case of a 
dipole of given size $R$ scattering on a dipole of energy-dependent size $R(\s)\sim R\s^{-\f \lam 2},$
one finds $L_{tail}\sim R\s^{\f 27-\lam \f 3{14}}.$ This has the expected effect of
enlarging the domain of elasticity and, as we shall see now, to strengthen the 
high-energy bound on the total cross section.

\subsection{A convergence problem for $\Re\, {\cal A }_{tail}$}

As a preliminary, we have to discuss the convergence properties of
${\cal A }_{tail}$ in \eqref{Tail}. Performing the angular
integration, one has (with $q\equiv|\vec{q}\,|$)
\begin{equation}
  {\cal A}_{tail}(s,t;\vec{R}_1,\vec{R}_2) = -4\pi is\int_{L_{tail}}^{\infty} dL\,L\,
  J_0(qL)\, {\cal C}_M(\chi,L;R_1,R_2)\ .
\end{equation}
Following expression \eqref{eq:phaseshiftsM}, at large $L$ the integrand is 
dominated by the tachyonic KK scalar 
exchange\footnote{We have checked that the potentially divergent contribution coming
from the subleading (in energy) graviton terms (see Eqs.~(50-56)
in Ref.\cite{Jani}) actually cancel.} and behaves as 
\begin{equation}
  L \sqrt{\frac{2}{\pi q L}}\cos\left(qL-\frac{\pi}{4}\right) \left[
  i\left(\kappa_S\frac{R_1 R_2}{L^2}\frac{1}{\sinh\chi}\right) -
  \frac{1}{2}\left(\kappa_S\frac{R_1 R_2}{L^2}\frac{1}{\sinh\chi}\right)^2\right],
\end{equation}
and since the imaginary part is bounded by $L^{-3/2}$, while the real part is bounded
by $L^{-7/2}$,  the integral is
convergent as long as $q\neq 0$. 

However, if one sets $q=0$ directly in the integrand,
since $J_0(0)=1$ one finds a logarithmic divergence in the {\it real part}
of the amplitude, coming from this KK scalar contribution.
One can easily isolate the divergent part by writing
\eq
\begin{aligned}
  {\cal A}_{tail}(s,t;\vec{R}_1,\vec{R}_2)= &
- 4\pi is\int_{L_{tail}}^{\infty} dL\,L\, J_0(qL)\,\times\\
&\times\left[\f{R_1R_2}{L^2}\f{i\kappa_S}{\sinh\chi}+ \left({\cal
    C}_M(\chi,L;R_1,R_2)
  -\f{R_1R_2}{L^2}\f{i\kappa_S}{\sinh\chi}\right)\right]\,.
\end{aligned}
\eqx
The term in brackets is convergent at $q=0$; to treat the other term
one divides the integral as follows:
\eq
\begin{aligned}
  {\cal A}_{tail}^{div}(s,t;\vec{R}_1,\vec{R}_2) \equiv& - 4\pi is\int_{L_{tail}}^{\infty} dL\,L\, J_0(qL)\,
 \f{R_1R_2}{L^2}\f{i\kappa_S}{\sinh\chi}
\\ \, 
=& \,4\pi \kappa_S \f{s}{\sinh\chi}R_1R_2\left[\int_{1}^{\infty}
\f{d\lambda}{\lambda}\, J_0(\lambda)+
\int_{qL_{tail}}^{1} \f{d\lambda}{\lambda}\, (J_0(\lambda) -1)\, + \right. \\ & \left. \hphantom{4\pi \kappa_S \f{s}{\sinh\chi}R_1R_2\bigg[ } +
\int_{qL_{tail}}^{1}\f{d\lambda}{\lambda}\right].
\end{aligned}
\eqx
It is now easy to see that the only divergent term when $q\to 0$ is the last one,
and so we conclude
\begin{equation}
  {\cal A}_{tail}^{div}(s,t\to 0;\vec{R}_1,\vec{R}_2) = 4\pi \kappa_S
  \frac{s}{\sinh\chi} R_1 R_2\log\frac{1}{qL_{tail}} + {\rm finite\,\,terms}\ ;
\end{equation}
recalling the relation \eqref{chi} between $\chi$ and $s$ one then obtains in the
high-energy limit
\begin{equation}
  {\cal A}_{tail}^{div}(s,t\to 0;\vec{R}_1,\vec{R}_2) = 8\pi
  \kappa_S m_1m_2 R_1R_2\log\frac{1}{qL_{tail}}  + {\rm finite\,\,terms}\ .
\end{equation}
In fact, one should distinguish the {\it mathematical} problem of determining 
the real part of the amplitude from a deeper {\it physical} one concerning
the AdS/CFT correspondence itself. Indeed,
since the imaginary part of the amplitude is always
finite in the $t\to 0$ limit, and moreover, as we will see in a
moment, also analytic in $s$, it is known how to obtain the real part
by means of a  dispersion relation, which yields a
finite result. However, on the gravity side, it is as yet unclear what is
the origin of this divergence, that we  expect to be cancelled by
effects which do not show up at the given level of supergravity approximation
of the AdS/CFT correspondence.
Moreover, for our purpose, the dependence on the energy is weak, coming through the
dependence of $L_{tail}$ on $s$. In the following we will then discard
this divergence, focusing on the dominant contribution at large $s$; conceptually,
it may be a relevant issue, but we delay this study for the future.

\section{The  $\nnn=4$ SYM forward amplitude}
\label{forward}
\subsection{Total cross section}
\label{fwdtcs}
We start from the imaginary part of the amplitude at $t=0$, which is
related to the dipole-dipole elastic total cross section by means of the optical
theorem. The contribution $\sigma_{tail}$ to the total cross section of the large
impact-parameter region as obtained from AdS/CFT  is given by
\eq
\begin{aligned}
  \sigma_{tail}& {\mathop \simeq_{s\to\infty}}  \f{\Im \, {\cal A}_{tail}(s,0;\vec{R}_1,\vec{R}_2)}{s} =
-4\pi \int_{L_{tail}}^{\infty}\! \!dL\,L\ \Re\, {\cal C}_M(\chi,L;R_1,R_2)  \\&=
  4\pi\int_{L_{tail}}^{\infty}\!\!
  dL\,L\, \left[1-\cos\left(\sum_\psi \dl_\psi\right)\right].
\label{total}
\end{aligned}
\eqx
The $\chi$-dependence  at large energy induces a
hierarchy  between the different contributions. 
This hierarchy  is clearly revealed after performing the change of
variables  
\ba
L\to \lam \equiv  (\sinh\chi)^{-\f{1}{6}} \f L{\sqrt{R_1R_2}};\quad L_{tail}\to 
\lam_{tail}=(\sinh\chi)^{-\f{1}{6}} \frac{L_{tail}}{\sqrt{R_1R_2}} \ ,
\label{change2}
\ea
which yields (rescaling with $\sinh\chi$ instead of
$\cosh\chi$ allows to keep manifest the symmetry under crossing, $i.e.$, under 
$\chi\to i\pi-\chi$, of the various phase shifts in formulas \eqref{eq:phaseshiftsM})
\eq 
\begin{aligned}
  \sigma_{tail} =& 4\pi(\sinh\chi)^{\f{1}{3}} R_1R_2
 \int_{\lam_{tail}}^{\infty} \! d\lam\,\lam \,
 \Bigg[1-\cos\Bigg(\f{\ka_S}{\lam^{2}}\frac{1}{(\sinh\chi)^{\f{4}{3}}} +
 \f{\ka_D}{\lam^{6}}\frac{1}{(\sinh\chi)^2}\, +\\ & \hphantom{ 4\pi(\sinh\chi)^{\f{1}{3}} R_1R_2
 \int_{\lam_{tail}}^{\infty} \! d\lam\,\lam \,
 \Bigg[} + 
   \f{\ka_B}{\lam^{4}}\frac{\coth\chi}{(\sinh\chi)^{\f{2}{3}}} + 
\f{\ka_G}{\lam^{6}}(\coth\chi)^2\Bigg)\Bigg]\, .
\label{integrallam}
\end{aligned}
\eqx 
The integral is clearly finite: 
\begin{itemize}
\item[-] the quantity in braces  
is always bounded between 0 and 2, which indeed corresponds to the 
unitarity constraint on the impact-parameter amplitude, 
\item[-] in the large $\lam$ limit the integrand behaves as $\lam^{-3}$
(corresponding to the KK scalar contribution to $\Im \, {\cal A}_{tail}(s,0)$).
\end{itemize}
The overall convergence is then ensured, contrary to the case of the real part.
Note that the leading term (the last one in \eqref{integrallam}) is
crossing-symmetric, thus  
corresponding to ``Pomeron exchange'' in the $S$-matrix language, while the
first subleading term, coming from 
antisymmetric-tensor exchange (the before-last one in \eqref{integrallam}),
 is crossing-antisymmetric under $\chi\to i\pi-\chi$, thus corresponding to
``Odderon exchange''. 
At large energy the dominant contribution is the one from graviton
exchange. Indeed, recalling the relation \eqref{chi} between $\chi$ and $s$ we obtain for $s\to\infty$
\eq
\begin{aligned}
 \sigma_{tail}  &\mathop \simeq_{s\to\infty} 4\pi {R}_{1}{R}_{2}
  \s^{\frac{1}{3}}\int_{\lam_{tail}}^{\infty} d\lambda \,\lambda
  \left[1-\cos\left(\frac{\ka_G}{\lambda^6}\right)\right] \\
&\mathop =_{\phantom{s\to\infty}} \f{2\pi}{3} {R}_{1}{R}_{2}\,\varsigma^{\f{1}{3}}
  \int_0^{\mu_{tail}} d\mu \,\mu^{-\f{4}{3}}
  \left[1-\cos\left(\ka_G\mu\right)\right]\\
  &\mathop =_{\phantom{s\to\infty}} \f{2\pi}{3} {R}_{1}{R}_{2}\,\left(\f{\s}{\mu_{tail}}\right)^{\f{1}{3}}
  \int_0^{1} dx\, x^{-\f{4}{3}}
  \left[1-\cos\left(\ka_G\mu_{tail}x\right)\right]\  ,
\label{mu}
\end{aligned}
\eqx
where we have set 
\begin{displaymath}
 \mu\equiv \lam^{-{6}}\ ;\quad \mu_{tail}= \lam^{-{6}}_{tail}\ ;\quad x=\f \mu{\mu_{tail}}\ . 
\end{displaymath}
To complete the
calculation of the high-energy behavior of $\sigma_{tail}$ we need
to know the limit of validity of the application of AdS/CFT and thus 
how $L_{tail}$ depends on $s$. Let us consider the parameterization 
\ba
L_{tail}=\lam_0 \sqrt{R_1R_2}
\ \varsigma^\beta\Rightarrow \lam_{tail}=\lam_0\ 
\varsigma^{\beta-\f{1}{6}}\ ;\quad\quad\mu_{tail}=\lam_0^{-6}
\varsigma^{1-6\beta}\ ,
\lab{para}
\ea
where $\lam_0$ may have some residual dependence on $R_{1,2}$ (see
$e.g.$~\eqref{e.constraint}).
 If $\beta<\f{1}{6}$ then
$\lam_{tail}\to 0$ and $\mu_{tail}\to\infty$ as the energy increases,
and so $\sigma_{tail} \sim \varsigma^{\f{1}{3}}$; if $\beta>\f{1}{6}$
then $\mu_{tail}\to 0$ and $\lam_{tail}\to\infty$, and so
$\sigma_{tail} \sim
\varsigma^{\f{1}{3}}\varsigma^{\f{5}{3}(1-6\beta)}=\varsigma^{2-10\beta}$. 
Finally, for $\beta=\f{1}{6}$ a constant $\lam_{tail}$ is found, and the
integral cannot modify the $s$-dependence. Summarizing, we have for large $s$
\eq
\begin{aligned}
  \label{eq:tail}
  \sigma_{tail} &\mathop \simeq_{\phantom{s\to \infty}} \ \f{2\pi}3\lam_0^2 R_1 R_2 \s^{2\bt}
\int_0^1  dx\, x^{-\f{4}{3}}\left[1-\cos \left(\ka_G \lam^{-6}_0 \s^{1-6\beta}x
\right)
\right] \\
&\mathop\sim_{s\to \infty} \f{2\pi}{3}R_1R_2\ \left\{
  \begin{aligned}
    &\s^{\f{1}{3}}\f{3\pi\ka_G^{\f{1}{3}}}{\Gamma(1/3)} & &\beta < \f{1}{6}\ ,\\
    &\s^{\f{1}{3}} \lam_0^2  \int_0^{1} dx\, x^{-\f{4}{3}}
    \left[1-\cos\left(\ka_G\lam_0^{-6}x\right)\right]     & & \beta=\f{1}{6}\ ,\\
    & \s^{2-10\beta}\frac{1}{2}\ka_G^2\lam_0^{-10} & & \beta>\f{1}{6}\ .
  \end{aligned}\right.
\end{aligned}
\eqx
We are now in the position to determine a lower and an upper bound on the
high-energy behavior of the dipole-dipole total cross section. 
Since obviously $\sigma_{tot}>\sigma_{tail}$, Eq.~\eqref{eq:tail} provides a
{\it lower} bound. The overall unitarity constraint Eq.~\eqref{Unit} 
allows
one to put an {\it upper} bound on the contribution from the $core$ region
$L<L_{tail}$, $i.e.$, $\sigma_{core} \le 4\pi L_{tail}^2 =
4\pi\lam_0^2R_1R_2\s^{2\beta}$, and thus on the whole total cross section,
\ba
\sigma_{tot}=\sigma_{core}\!+\sigma_{tail} <  4\pi\lam_0^2 R_1R_2 \s^{2\bt}
\left\{1+ \f{1}{6}\int_0^1  dx\, x^{-\f{4}{3}}\left[1-\cos \left(\ka_G \lam^{-6}_0 \s^{1-6\beta}x
\right)
\right]\right\}\! \, .
\lab{max}
\ea
We have included here only the leading part of the $tail$ contribution at high energy. 
A more rigorous way to write the $\beta$-dependent bounds on the
high-energy behavior of $\sigma_{tot}$ is the following,
\begin{equation}
  \label{eq:bound}
  \min\left(\f{1}{3},2-10\beta\right) \le \lim_{\s\to\infty}\f{\log \sigma_{tot}}{\log\s} \le
  \max\left(\f{1}{3},2\beta\right),
\end{equation}
which in particular, using the value $\beta=\f{2}{7}$ coming from the weak field
constraint \eqref{e.constraint}, yields the rigorous bound
\begin{equation}
  \label{eq:rigbound}
  -\f{6}{7} \le \lim_{\s\to\infty} \f{\log \sigma_{tot}}{\log\s} \le
  \f{4}{7}\,. 
\end{equation}
The following remarks are in order.
\begin{enumerate}
\item For $\beta<\f{1}{6}$, at sufficiently high energy one would 
have $L_{tail}<L_{min}$, thus entering the unphysical region where the impact-parameter 
partial amplitude is infinitely oscillating between 0 and 1. In this case the total cross 
section would become purely elastic at high energy, while one expects the opening 
of more and more inelastic channels as the energy increases: this means that one 
lies beyond the applicability of the elastic eikonal approximation. 
\item At $\beta=\f{1}{6},$ corresponding to $L_{tail}/L_{min}=const.$,  the $tail$ and $core$ contributions have the same
  high-energy behavior. In this case $\lam_{tail}$ (or, equivalently,
  $\mu_{tail}$) does not depend on energy. However, one has to verify the 
  non-oscillating behavior condition
  \ba
  \lam_{tail}\ge \left(\f {\ka_G}\pi\right)^{\f16},\quad\quad
 \mu_{tail}\le  \f \pi{\ka_G}\ .
  \lab{Osci}
  \ea
\item For $\f16<\bt\le\f{2}{7}$, which corresponds to $L_{min}<L_{tail}<L_{max}$ (strictly speaking, at sufficiently high energy), 
  the $core$ region dominates, while the $tail$ region gives a
  subleading contribution as $s\to\infty$. The two bounds determine a
  window of possible power-law behaviors. 
\item For the maximal value  $\beta= \f 27$, $i.e.$, for $L_{tail}/L_{max}=const.$, the total
  cross section behavior is constrained to be such that $\sigma_{tot}\le { const.}\times\s^{\f 47}.$ 
  This maximal value is determined from the requirement that the AdS/CFT correspondence can be
  reliably applied, $i.e.$, that the constraint \eqref{limit} for the gravitational perturbation to be weak
  is verified. In fact, this is the rigorous result obtained 
  by means of the AdS/CFT correspondence, since for smaller $\bt$ one expects inelastic contributions
  coming from a strong dual gravitational field.
\item One could also consider $\bt>\f 27$, but in that case one would only obtain a weaker bound on the total cross section. Indeed, in doing so one
  would overestimate the contribution of the $core$, including in it the impact-parameter region $L_{max}<L<L_{tail}$, where the amplitude is reliably
  described by the eikonal AdS/CFT expression.
\end{enumerate}
In the rest of this paper we will thus consider the $\bt$-dependence in the domain
$\f16\le\bt\le\f{2}{7}$. The limiting values have the following
characteristics: $\bt=\f{1}{6}$ is the minimal admissible value for which the
eikonal approximation using real phase shifts could be valid, and $\f{2}{7}$
is the absolute bound coming from the AdS/CFT correspondence. 
Note that from Eq.~\eqref{eq:tail} we  see that  $\Im \,{\cal
  A}_{tail}(s,0)=\varsigma^{1+\g}\times (s\text{-independent})$ is
analytic in $s$, with a branch point at $s=0$. 
\FIGURE{
  \centering
  \includegraphics[width=0.8\textwidth]{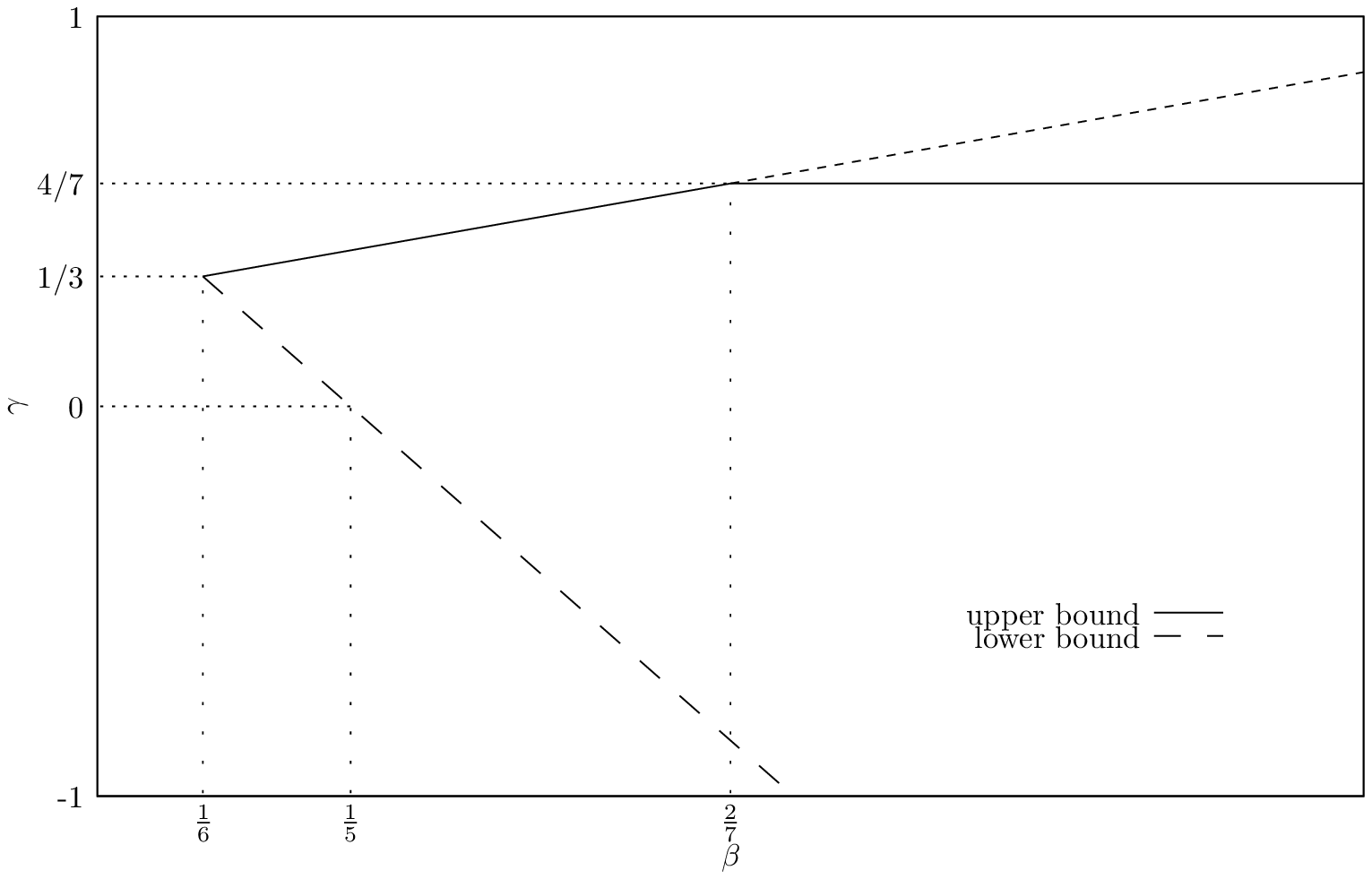}
\caption{{\it Upper and lower bounds on the high-energy behavior of
    total cross  sections.} The bounds on the total cross section exponent $\gamma$,
  $i.e.$, on the Pomeron intercept minus one, are displayed as a function of the power law exponent $\bt$ of 
  $L_{tail}\propto \s^{\bt}$. {\it Solid line}: upper bound, coming from
  the $core$ contribution for $\f{1}{6}\le\bt\le\f 27$.  {\it Long-dashed line}: lower
  bound, coming from the $tail$ contribution for $\bt>\f{1}{6}$. 
{\it Short-dashed line}: weaker upper bound for $\bt > \f 27$, obtained by overestimating the $core$ contribution (see text). 
  We singularize the value $\bt=\f{1}{5}$ below which  the $tail$ contribution and thus the total cross section are bounded to grow.}
}

\subsection{Asymptotic forward phase}
A general result regarding scattering amplitudes at high-energy
relates the phase $\varphi$ of the amplitude to the leading behavior
in $s$ (see $e.g.$~\cite{Eden,Morel} and references therein):
for a symmetric amplitude behaving as $\sim s^\alpha$ at 
high energy one has ${\cal A}^{(+)} \sim \pm
s^\alpha e^{i\pi(1-\frac{\alpha}{2})}$, while for an antisymmetric
amplitude with the same leading $s$-dependence one finds ${\cal
  A}^{(-)} \sim \pm s^\alpha
e^{i\pi(\frac{1}{2}-\frac{\alpha}{2})}$. Note that for the dominant 
contribution the sign ambiguity is fixed by 
asking for a positive total cross section. This result is a consequence
of analyticity, and it is obtained through the application of the 
Phragm\'en-Lindel\"of theorem to the function ${\cal
  A}^{(\pm)}/(s-u)^\alpha$ ($u$ is here the usual Mandelstam variable). Hence we have at asymptotic energy
\ba
\f{\Im\,{\cal A}^{(+)}}{\Re\,{\cal A}^{(+)}}= -\tan \f{\pi\al}2\ ;\quad\quad
\f{\Im\,{\cal A}^{(-)}}{\Re\,{\cal A}^{(-)}}= \cot \f{\pi\al}2\ .
\lab{ImRe}
\ea
Although this result holds in general for any value of $t$ along the
``Regge trajectory'' $\al(t)$, we consider here the forward
amplitude.
Using \eqref{ImRe} for a positive signature amplitude ${\cal A}^{(+)}$ 
with $\al=1+\g$, and taking $\g=\g(\bt)$ to be the upper bound on the exponent 
determined above in section \ref{fwdtcs}, we find that at large $s$ 
\begin{equation}
  \label{Forw}
  {\cal A}^{(+)} = C\s^{1+\g}e^{i\f{\pi}{2}(1-\g)}, \quad \g(\bt)=\max\left(\f{1}{3},2\beta\right)\ ,
\end{equation}
where the phase $\varphi=\f{\pi}{2}(1-\g)$ varies in the range between
$\varphi=\f{3\pi}{14}$, for the maximally $core$-dominated $\bt= \f 27$ 
amplitude, and $\varphi=\f{\pi}{3}$ for the $tail$-dominated result at $\bt\le
\f 16$. Here $C$ is a positive constant, and we fix
the sign ambiguity of the amplitude by requiring the positivity of the
total cross section. 

The interesting outcome of these analyticity properties is that
we gain a new constraint on the overall amplitude, which can help 
complementing the knowledge of the $tail$ region from AdS/CFT. As 
an example, one may consider a ``black disk model'' \cite{Black}, 
where one assumes the eikonal approximation in the whole $tail$ ($i.e.$, $\bt=\f16$), 
and  a maximally inelastic (``black disk'') amplitude in the $core$ 
($i.e.$, $a(\chi,\vec b)\sim i$). One then obtains the frontier between 
$tail$ and $core$ being fixed at $\kappa_G\mu_{tail}\sim \f\pi 2$ 
with a forward  amplitude consistent with analyticity and  unitarity 
and satisfying the constraint relation \eqref{Osci}.

\newpage

\subsection{Subleading contributions}

The expression Eq.~\eqref{integrallam} shows clearly the hierarchy in
energy of the contributions of the various supergravity fields. Indeed,
at large $s$, keeping only the leading contribution from each field,
we have for the subleading part of the total cross section
\ba 
\sigma^{subleading}_{tail} \simeq 4\pi\s^{\f{1}{3}} R_1R_2
 \int_{\lam_{tail}}^{\infty}  d\lam\,\lam \,
 \sin\left(\f{\ka_G}{\lam^{6}}\right)\left(\f{\ka_S}{\lam^{2}}\frac{1}{\varsigma^{\f{4}{3}}}
+ \f{\ka_D}{\lam^{6}}\frac{1}{\varsigma^2} + 
   \f{\ka_B}{\lam^{4}}\frac{1}{\varsigma^{\f{2}{3}}}\right) \ .
\lab{Absorb}
\ea 
The well-known  ``absorption''  phenomenon  appears in \eqref{Absorb}, 
since the secondary contributions are shielded by the 
$\sin\left(\f{\ka_G}{\lam^{6}}\right)$ coming from the 
leading graviton contribution. Its natural interpretation in a $S$-matrix framework comes from 
the initial and final state elastic interactions which 
correct the ``bare'' secondary contributions.

For $\beta> \f 16$, in which case $\lam^{-1}_{tail}\to 0$ in \eqref{Absorb}, we have for the $tail$ contributions
\begin{align}
  &\begin{aligned}
    \lab{Hier2}
    &\text{graviton \phantom{({\it tail}\ \rm{and}\ {\it core}\/)}} & &\lra &  &2-10\beta\\
    &\text{antisymmetric tensor} & &\lra & &1-8\beta\\
    &\text{KK scalar} & &\lra & & \phantom{1}-6\beta\\
    &\text{dilaton} & &\lra & & \phantom{1}-10\beta\ .
  \end{aligned}
  \intertext{The secondary contributions in the $core$ are neither determined 
    nor usefully constrained by unitarity. 
    Note that, in the limiting case $\bt=\f 16$, with $\lam_{tail}\sim const.$,  the leading
    $s$-dependence of the various $tail$ contributions to the total cross section
    obey the following hierarchy:}
  &\begin{aligned}
    \lab{Hier}
    &\text{graviton ({\it tail}\ \rm{and}\ {\it core}\/)} & &\lra &  &\textstyle\f{1}{3}\\
    &\text{antisymmetric tensor} &&\lra & -& \textstyle\f{1}{3}\\
    &\text{KK scalar} & &\lra & -& 1\phantom{\f{1}{1}}\\
    &\text{dilaton} & &\lra & -& \textstyle\f{5}{3}\ .
  \end{aligned}
\end{align}
Starting from the graviton, the intercept of the other contributions is
obtained subtracting $2/3$ each time. Note that, when expanding the eikonal
expression, one must be aware that subleading contributions from a
field with a larger intercept mix with the leading contributions of
the less relevant fields. 

In the ``black disk model'' \cite{Black}, one assumes total absorption  in the $core$ region, an thus formula \eqref{Absorb} 
would give the whole contribution to the total cross section from  subleading contributions. 
Hence, in that case one would determine  the effective hierarchy
of intercepts to be given by \eqref{Hier} with $\bt=\f 16$. In particular, the
``Odderon'' contribution, $via$ Gauge/Gravity duality, is found to have intercept $\f23$, 
which is less than the perturbative $1$ corresponding to the
exchange of 3 gluons. 

\section{Summary, discussion and outlook}
\label{concl}
Using the AdS/CFT correspondence to determine the dipole-dipole elastic amplitudes
at large impact-parameter, and the constraints from analyticity and unitarity at 
lower impact-parameter, we study the high-energy behavior of soft amplitudes in
$\nnn=4$  SYM gauge field theory. Our results can be summarized as follows.
\en
\item
In the region where the AdS/CFT correspondence is fully valid, $i.e.$, 
at $L>L_{max}$ where the supergravity field is weak enough, we found an absolute bound 
$\sigma_{tot}< const. \times \s^{\f 47}$ for the high-energy behavior of the total
cross section. In the usual language of strong interactions it corresponds 
to the bound  $\f {11}7$ for the ``Pomeron intercept''.
This bound is governed by the graviton exchange in the dual AdS bulk.
\item
Below $L_{max}$, there are relevant regions in the bulk where the induced gravitational field becomes strong w.r.t. 
the background AdS metric. Hence, one would expect self-interacting graviton exchanges, which could spoil the elastic eikonal expression. 

\item
The upper 
bound is strengthened to give $\sigma_{tot}< const. \times \s^{2\bt}$, if one adopts the hypothesis of validity of the eikonal 
approximation in the impact-parameter region $L>L_{tail}\propto \s^\bt$, with $\f 16 <\bt<\f 27$, thus using independent graviton exchange even when strong  
gravitational perturbations appear in the bulk.

\item
The eikonal approximation with independent graviton exchange cannot be valid below  an 
impact-parameter $L_{min} \sim \s^{\f 16},$ from the physically motivated
requirement of non-oscillating cross sections. For this value, the
$core$ and the $tail$ contributions to the impact-parameter amplitude
have the same power in energy, transforming the bounds
into a prediction $\sigma_{tot}\sim  \s^{\f 13}$, $i.e.$, a ``Pomeron
intercept'' equal to $\f 43$.
\item 
The real part of the forward amplitude coming from the AdS/CFT
determination contains a divergence which can be got rid of using
dispersion relations. However, it points towards a necessary
completion of the AdS/CFT correspondence beyond the exchange of the
tachyonic KK scalar mode. 
\enx
In order to obtain these results, we made use of the minimal surface
formulation of the AdS/CFT correspondence of Ref.~\cite{Jani}. The
elastic  amplitudes at large impact-parameter are combined with analyticity and unitarity constraints to
evaluate the behavior of the total cross sections. It is useful to compare our results obtained using this
method with the other existing approaches. 

The absolute bound we obtain is a new result, which is linked to the
precise derivation of a weak gravitational field limitation of the
AdS/CFT correspondence in the supergravity formulation. We note that
the upper bound $\sigma_{tot}<const. \times \s^{\f 47}$ necessarily
restricts the total cross section to be below the $bare$ 
graviton exchange contribution, namely $\sigma_{tot}\sim  \s^{1}.$ 

Our result appears as the analogue of the Froissart bound, but in
the context of the non confining $\nnn=4$ SYM theory, since it is the
combination of the unitarity bound on impact-parameter amplitudes with
the determination of a precise power-like bound on the
impact-parameter radius from AdS/CFT (for confining theories, this
bound is logarithmic leading to the Froissart bound $\sigma_{tot}\sim
\log^{2}\s$). 

We remark that a more stringent bound would be obtained if one assumed
the validity of the elastic eikonal approximation in a region with
strong gravitational field in the bulk. This is why we considered the
possibility of an enlarged impact-parameter region of validity of the
eikonal approximation, defined by a power-like behavior $L >
L_{tail}\propto \s^\bt$, with $\f 16 <\bt<\f 
27$. This results into a $\bt$-dependent bound. However, in the region
of strong gravitational fields, as we have discussed in section \ref{forward},
other contributions are expected to modify the gravitational sector.  

Contributions to the graviton Reggeization
\cite{Brow0,Brow1,Brow2,Brow3} have not been considered in the present
calculations. These are corrections  which are beyond the
supergravity approximation. Indeed, for the bare Pomeron propagator,
these can be justified from a valid flat space approximation
\cite{Brow0,Polc}. However, we do not know yet how to couple in a
consistent way this improved propagator to the disconnected minimal
surfaces at large impact-parameter. This is an interesting subject to
be studied further on. 

The remarkable value $\f 43$ of the Pomeron intercept common for the $tail$ and 
the $core$ contributions  in the special
case $L_{tail}\sim \s^{\f 16}$ has been already previously noticed  using
minimal surfaces \cite{Matt}. This result has been obtained recently without 
using minimal surfaces in \cite{LP}, where the authors 
also notice  that the reggeized correction should appear at much
higher energies than the initial intercept $\f 43.$ It would be informative to understand the 
relation between minimal surfaces and their derivation.
Note that in a subsequent
paper \cite{Khar}, a new source of inelasticity is discussed beyond the AdS/CFT correspondence. 
Indeed, the study of a convenient description of the central impact-parameter region with a strong
inelastic  component is an important open topic. 

There exist related AdS/CFT calculations of a Wilson loop immersed
into a gauge field background, whose dual description is a modified
AdS metric, and which aims at describing DIS on a large nucleus
\cite{Tali1,Tali2}. Indeed, these studies look for a determination of
the Pomeron intercept which could be compared to our bounds. For
instance, in \cite{Tali1} two solutions have been found with intercept
$2$ and $\f32,$ the first one violating the black disk limit
\cite{Tali2}. The second one is in agreement with our absolute bound. As
an outlook, it would be interesting to know the impact-parameter dependence of this solution.

Finally, it would be interesting to extend this study to the overall
elastic amplitude, by using more constraints, $e.g.$ the analyticity
requirements at non zero momentum transfer. Also, the physically
interesting case of confining theories could be studied in more
details using as an input the results of the minimal surface studies of
Refs.~\cite{Jani1,Jani2}. 

\acknowledgments
R.~P.~wants to acknowledge the long-term  collaboration
with Romuald Janik, whose results are used in the present paper. We
warmly thank Cyrille Marquet for a careful reading of the manuscript
and useful suggestions, and Enrico Meggiolaro and Fernando Navarra for
fruitful remarks. M.~G.~thanks the Institut de Physique Th\'eorique,
Saclay, for the scientific invitation. This work has been partly
funded by a grant of the ``Fondazione Angelo Della Riccia'' (Firenze,
Italy).

\end{document}